\documentclass[preprint]{aastex}
\usepackage{psfig}
\usepackage{natbib}

\newcommand{\WMAP}{{\sl WMAP}}
\newcommand{\Barnes}{{C. Barnes}}
\newcommand{\Bennett}{{C. L. Bennett}}

\newcommand{\Halpern}{{M. Halpern}}
\newcommand{\Hill}{{R. S. Hill}}
\newcommand{\Hinshaw}{{G. Hinshaw}}
\newcommand{\Jarosik}{{N. Jarosik}}
\newcommand{\Kogut}{{A. Kogut}}

\newcommand{\Limon}{{M. Limon}}
\newcommand{\Meyer}{{S. S. Meyer}}

\newcommand{\Page}{{L. Page}}

\newcommand{\Tucker}{{G. S. Tucker}}

\newcommand{\Wollack}{{E. Wollack}}
\newcommand{\Wright}{{E. L. Wright}}
\newcommand{\Brown}{{Dept. of Physics, Brown University, %
            Providence, RI 02912}}
\newcommand{\Goddard}{{Code 685, Goddard Space Flight Center, %
            Greenbelt, MD 20771}}
\newcommand{\NRCFellow}{{National Research Council (NRC) Fellow}}
\newcommand{\PrincetonPhysics}{{Dept. of Physics, Jadwin Hall, %
            Princeton, NJ 08544}}
\newcommand{\PrincetonAstro}{{Dept of Astrophysical Sciences, %
            Princeton University, Princeton, NJ 08544}}
\newcommand{\SSAI}{{Science Systems and Applications, Inc. (SSAI), %
            10210 Greenbelt Road, Suite 600 Lanham, Maryland 20706}}
\newcommand{\UBC}{{Dept. of Physics and Astronomy, University of %
            British Columbia, Vancouver, BC  Canada V6T 1Z1}}
\newcommand{\UChicago}{{Depts. of Astrophysics and Physics, EFI and CfCP, %
            University of Chicago, Chicago, IL 60637}}
\newcommand{\UCLA}{{UCLA Astronomy, PO Box 951562, Los Angeles, CA 90095-1562}}

\renewcommand{\WMAP}{\mbox{\sl WMAP~}}

\renewcommand{\deg}{\mbox{$^\circ$}}

\slugcomment{Submitted to {\it The Astrophysical Journal}}

\newcommand{\dg}           {\mbox{$^{\circ}$}}
\newcommand{\lsim}         {\mbox{$_<\atop^{\sim}$}}

\newcommand{\muK}          {\hbox{$\mu$K}}

\newcommand{\be}           {\begin{equation}}
\newcommand{\ee}           {\end{equation}}
\newcommand{\ba}           {\begin{eqnarray}}
\newcommand{\ea}           {\end{eqnarray}}
\newcommand{\labeq}[1] {\label{eq:#1}}
\newcommand{\eqn}[1] {(\ref{eq:#1})}
\newcommand{\labfig}[1] {\label{fig:#1}}
\newcommand{\fig}[1] {\ref{fig:#1}}

\newcommand{\mb}[1] {\mbox{\boldmath $#1$}}
\newcommand{\dee}      {\,{\rm d}}

\begin{document}

\title{First Year {\sl Wilkinson Microwave Anisotropy Probe} (\WMAP) 
Observations:
Galactic Signal Contamination from Sidelobe Pickup}

\author{
\Barnes\altaffilmark{2}, 
\Hill\altaffilmark{9},
\Hinshaw \altaffilmark{3}, 
\Page\altaffilmark{2},  
\Bennett\altaffilmark{3},
\Halpern \altaffilmark{4}, 
\Jarosik \altaffilmark{1}, 
\Kogut \altaffilmark{3}, 
\Limon \altaffilmark{2,3,10}, 
\Meyer \altaffilmark{5},
% \Spergel \altaffilmark{6}, 
\Tucker \altaffilmark{7,10},
\Wollack \altaffilmark{3},
\Wright \altaffilmark{8}
}

\altaffiltext{1}{\WMAP is the result of a partnership between Princeton 
                 University and NASA's Goddard Space Flight Center. Scientific 
		 guidance is provided by the \WMAP Science Team.}
\altaffiltext{2}{\PrincetonPhysics}
\altaffiltext{3}{\Goddard}
\altaffiltext{4}{\UBC}                 
\altaffiltext{5}{\UChicago}
\altaffiltext{6}{\PrincetonAstro}
\altaffiltext{7}{\Brown}
\altaffiltext{8}{\UCLA}
\altaffiltext{9}{\SSAI}
\altaffiltext{10}{\NRCFellow} 

\email{cbarnes@princeton.edu}


\begin{abstract}
Since the Galactic center is $\sim 1000$ times brighter than
fluctuations in the Cosmic Microwave Background (CMB), CMB
experiments must carefully account for stray Galactic pickup.  
We present the level of contamination due to sidelobes for the
year one CMB maps produced by the \WMAP observatory. For each radiometer,
full $4\pi$ sr antenna gain patterns are determined from a combination
of numerical prediction, ground-based and space-based measurements.
These patterns are convolved with the {\sl WMAP} year one sky
maps and observatory scan pattern to generate expected sidelobe signal
contamination, for both intensity and polarized microwave sky maps.
Outside of the Galactic plane, we find rms values for the  expected sidelobe
pickup of 15, 2.1, 2.0, 0.3, 0.5~$\muK$\ for K, Ka, Q, V, and W-bands
respectively.  Except at K-band, the rms polarized contamination is $\ll
1\muK$. Angular power spectra of the Galactic pickup are presented.

\end{abstract}

\section{Introduction}

{\sl WMAP} consists of dual back-to-back Gregorian telescopes
designed to differentially measure fluctuations in the Cosmic
Microwave Background (CMB) \citep{bennett/etal:2003}. 
\WMAP  is designed to create maps of the microwave sky in five 
frequency bands, generically labeled K, Ka, Q, V, and W bands,
centered on 23, 33, 41, 61, and 94 GHz respectively. Like all radio
telescopes, each \WMAP beam has sidelobes, regions of nonzero gain
away from the peak line-of-sight direction.

The brightest sidelobes for each \WMAP beam
correspond to radiation paths which, if started from the feeds, miss a
telescope reflector and
 go to regions of the sky far from the main beam.  
Figure~\fig{optics} shows a rendition of \WMAP's optical structure.
The most important sidelobes result from radiation which, if coming
from the sky, spills past the 
primary reflector, striking the secondary either directly 
or after bouncing off of one of the two radiator panels.  These light
paths create angularly broad, smooth swaths of antenna gain
20\deg to 100\deg\ from the beam peaks.   In addition,
multiple reflections between the secondary reflector and the front of the
Focal Plane Assembly (FPA) of feed horns contribute to a complicated
`pedestal' of gain surrounding each beam peak,  $\sim 2$\deg to
15\deg\  
from the beam peak position.  Radiation which follows the intended optical
path (primary reflector to secondary to feed horn) is considered part of
the main beam, and is considered in a companion paper 
\citep{page/etal:2003b}.

\begin{figure}
\epsscale{0.8}
\plotone{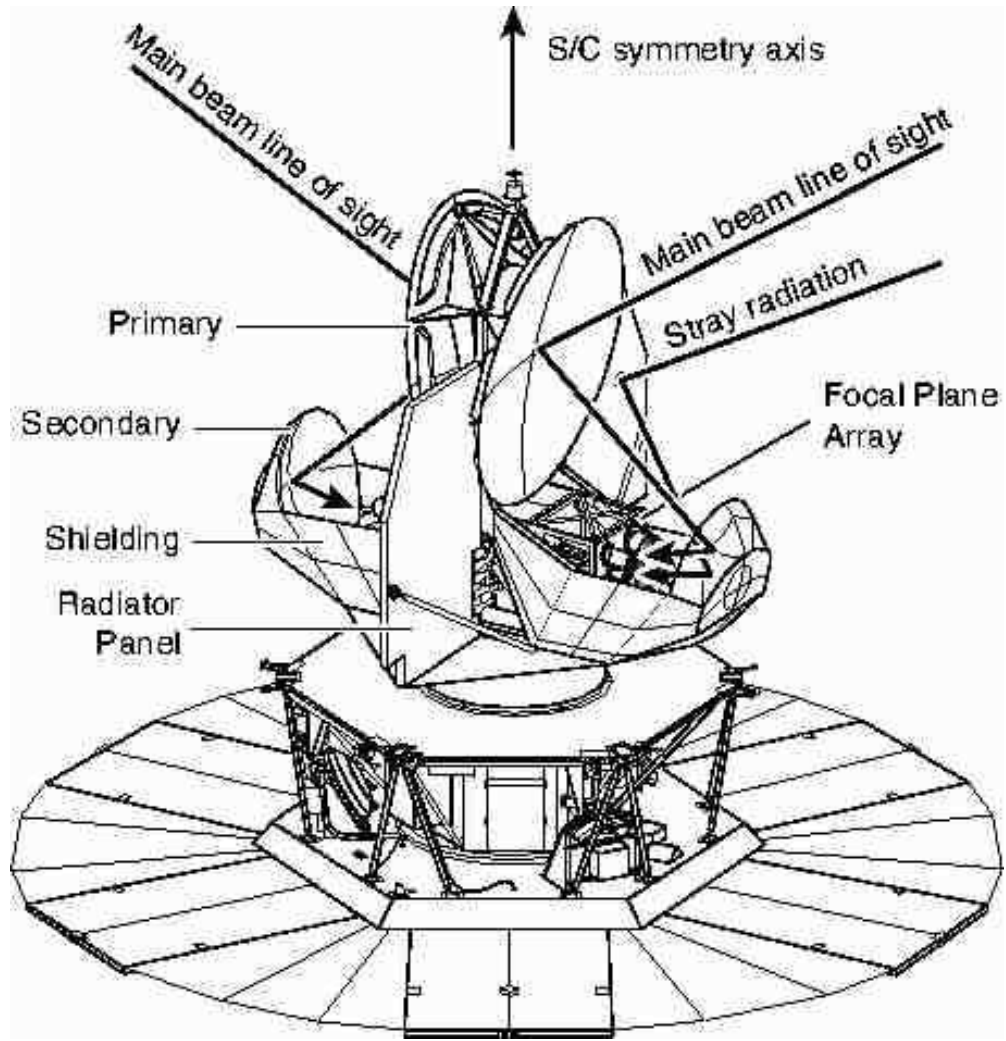}
\caption{ A line drawing of \WMAP spacecraft showing the principle optics
elements. Radiation from the sky at 68\deg\ from the spacecraft axis
of symmetry is reflected from the two back-to-back primary reflectors to
the two secondary reflectors. Diffraction shields half surround the
secondary reflectors to limit radiation from below. The secondaries
reflect the radiation into the arrays of feed horns near the center of
the instrument. The main diffraction paths are past the edges of the
primary reflectors, either directly from the sky or reflecting from the
flat radiator panels behind and between the primaries.
\textbf{\textsl{
A higher quality rendering is available on the LAMBDA web site.}}}
\labfig{optics}
\end{figure}

Sidelobe gains are most usefully expressed in dBi =
$10\log_{10}G(\theta,\phi)$, where the gain $G$ is normalized so that
a non-directional antenna has $G(\theta,\phi)=1$, or 0 dBi over $4\pi$
sr.  The maximum sidelobe gains range from 5 dBi (in K-band) to 0 dBi
(in W-band), from 40 dB to 60 dB below the beam peak gains.  For most
applications in radio astronomy, such weak responses would be
ignorable.  However, the relative brightness of Galactic foregrounds
makes sidelobe pickup a potentially significant systematic effect for
CMB measurements. For the \WMAP optics, sidelobe pickup amounts range
from $3.7\%$ to $0.5\%$ of total sky sensitivity, K-band to W-band
respectively \citep{page/etal:2003}. Sidelobe pickup introduces a
systematic additive signal into the time ordered data (TOD) for each \WMAP
radiometer differencing assembly (DA).  This signal propagates through
the map-making algorithm \citep{hinshaw/etal:2003} into the final sky
map. Since sidelobe pickup does not enter the data stream in the same
form as the desired sky signal, the overall contamination in the
derived microwave sky maps is a factor of 2 to 3 less than than the sidelobe
contribution to the original time ordered data.

For both polarized and unpolarized sky maps, we find that sidelobe
contamination is strongest at K-band, growing negligible toward V and
W-band.  For unpolarized sky maps, the rms sidelobe-induced signal per
pixel is $15\ \muK$ at K-band, and $2\ \muK$ or less for Ka, Q, V, and W-bands.
The sidelobe-induced polarized contamination is much weaker, $1\ \muK$
rms per pixel at K-band and $\le 100$ nK in all other bands.  (These
averages reflect the CMB analysis region of the sky only; specifically
the Kp0 cut \citep{bennett/etal:2003b}.)  In each case significant sidelobe
pickup is confined to the lowest spherical harmonics, $1\le \ell \lsim
10$.

This paper is organized as follows.  In 
\S~\ref{section:antenna_patterns}, we discuss the measurement,
physical optics models and calculation of the full sky antenna patterns.
In \S~\ref{section:calculation}, we describe how these 
sidelobe patterns are convolved with the one-year measured sky
patterns to calculate the Galactic signal contamination within 
each \WMAP first year sky map.  Contributions to unpolarized and 
polarized microwave maps are calculated, and compared with the CMB
signal. 

A set of sidelobe contamination maps are available with the main \WMAP
data release.

\section{Determination of the Sidelobe Gain Patterns}
\label{section:antenna_patterns}

Antenna gain patterns for \WMAP's optics were measured at two
antenna ranges on the ground, and by using the Moon as a bright source
during the spacecraft's phasing loops \citep{bennett/etal:2003}.  In addition,
full-sky antenna patterns were modeled using physical optics
software (YRS \cite{YRS:DADRA}).  No one measurement or model alone provides
enough information to construct an accurate, calibrated, full sky
antenna pattern; each has a different region of applicability. 
In combination, the methods provide a sufficient sidelobe map
for the analysis of the \WMAP data.

\subsection{Ground-based Measurements}

Ground-based sidelobe measurements were done at the Goddard
Electromagnetic Anechoic Chamber (GEMAC), which used a prototype of
one side of the \WMAP optics, and an outdoor test range at
Princeton University, which used a mockup of the complete spacecraft.
Both ranges used only single-frequency (narrow-band) measurements of
antenna response.  Since both the Galaxy and \WMAP radiometers have
broadband microwave response, single-frequency gain measurements cannot
be taken as the final effective antenna pattern.  In both cases, the
microwave sources are linearly polarized, permitting polarized
measurements of the sidelobe gain patterns.

\subsubsection{GEMAC Measurements} 

The GEMAC is an indoor range consisting of a 3m collimating mirror
within an anechoic chamber. The range allows absolute signal
calibration to 0.1 dB, with ambient reflection levels ranging from -40
dBi in K-band to -20 dBi in W-band.  The collimating mirror renders
the incoming wavefronts parallel, placing the source in the
telescope's far-field for all frequency bands.  The GEMAC uses an
azimuth-over-elevation type mount to rotate the telescope, allowing
precise ($0.005\deg$) pointing control.  This range cannot observe
gains for source elevations lower than $38\deg$ below the spacecraft
horizon, or higher than $43\deg$ above it.

Inside the GEMAC range, gain patterns were measured for the optical
assembly (feed horns, primary and secondary reflectors, and
supporting structure) alone, without radiator panels or solar shields
in place.  Consequently large sections of the far sidelobes---light
which reflects from the radiator panels in flight---wind up in the
wrong place.  Since the radiator panels are flat reflectors, to an
excellent approximation the GEMAC far sidelobe patterns are correct in
shape and intensity, though reflected and displaced on the sky.

\subsubsection{Princeton Measurements}

Princeton's antenna range has less accurate pointing
($0.1\deg$), and permits only relative signal calibration, but has a
significantly lower noise floor than the GEMAC range.  At the Princeton
test range, the telescope is mounted on a rooftop with an
elevation-over-azimuth type drive, while the microwave source is set
on a tower 91 m away.

The Princeton range is outdoors, and is arranged so the telescope's
main beam points at empty sky when it is not looking directly at the
source.  Any reflective pickup must bounce off the ground underneath
the telescope, and come in from a disfavored direction.  With one
exceptional region (described below), the noise floor of the Princeton
sidelobe-pattern measurements was determined by receiver noise.
Noise floors ranged from -50 dBi in K-band to -35 dBi in Q-band.  (V
and W-band measurements were limited at the Princeton range by the
weakness of the sidelobes, detector noise and uncertainty of
calibration.)

At 91 m, the source is in the near field of the telescope's primary,
which has an effective diameter of 1.4 m.  Beam peak measurements are
out of focus. However, for the purpose of measuring sidelobe response,
this condition is immaterial.  Sidelobes come from light paths which
miss the primary reflector.  For this unfocused light, the relevant
antenna size is the aperture of the feed horns. As the horn apertures
range from 4 to 11 cm in diameter, with the far-field beginning at
$4D^2/\lambda \sim 4$ m, a source 91 m away is well in the far field.
Consequently the Princeton antenna range is usable for measuring
antenna patterns everywhere except in the immediate neighborhoods of
the beam peaks.

At Princeton, the reflectors were mounted to a complete model
of the spacecraft; no important radiation paths should differ
materially from the observatory in flight.

\smallskip

In summary the Princeton test range data is usable everywhere away
from the main beam peaks, while the GEMAC data is usable everywhere 
except for reflections and shadows of the satellite radiator
panels. 

\subsection{In-flight Lunar Measurements}

The most reliable sidelobe measurements come from the spacecraft's
first orbital phasing loop, just after launch. Data using the Moon as
a microwave source were taken between July 2 and July 8 2001.  During
a total of 4.3 days within this period, \WMAP was in normal observing
mode, i.e., scanning the sky with a motion incorporating both spin and
precession.  From the spacecraft point of view, the angular diameter
of the Moon varied by a factor of $\sim 2$, with a mean of 1.2\deg.  
In spacecraft angular coordinates, the Moon swept out a path
that provided coverage of the upper hemisphere down to a latitude
of 25\deg, which is just above the main beams.

In-flight Moon measurements are preferred to ground
measurements because:

\smallskip
The Moon is a roughly thermal microwave source, so Lunar
radiation allows broadband measurement of each antenna pattern. 
We treat the Moon as unpolarized.

The spacecraft/source geometry is exactly as for our CMB
measurements.  The Moon gives our only direct measurement of the
{\em differential} sidelobe pickup for the two back-to-back
telescopes.  All the other measurements are made from a single telescope,
half of Figure~\fig{optics}.

The radiometers, data collection, and pointing uncertainty 
on board the observatory are all superior to those on the ground.
\smallskip

The major limitation to using Moon data is the incomplete sky
coverage.

\subsubsection{Lunar Data Sampling}

For each of 10 pairs of feed horns, the sidelobe map comes directly
from the time ordered output of the A-B differential radiometer, binned
according to the Moon's position in satellite coordinates.  The
antenna gain is extracted directly from each measurement:
\be
G_X({\bf n}) = {\Delta T^{i,X}_{\rm obs} - \Delta T^{i,X}_{\rm sky} \over
T_{\rm Moon}} 
            \left( {4 \pi {R^i_{\rm M-S}}^2 \over \pi {R_{\rm Moon}}^2}\right)\ ,
\ee
where ${\bf n}$ is the Moon direction in satellite coordinates,
$T_{\rm obs}$ is the observed differential temperature, $R_{\rm M-S}$
is the Moon to satellite distance. The superscripts $X$ and $i$ denote
a particular horn position (e.g., Q2) and satellite
location/orientation respectively.  $\Delta T_{\rm sky}$ is the
differential sky signal (CMB + foregrounds) appropriate to the chosen
radiometer and observatory orientation.  $\Delta T_{\rm sky}$ is calculated
from the microwave sky maps.  Lastly, $T_{\rm Moon}$ is an effective
Lunar temperature for the microwave band in question, and carries the
gain calibration uncertainty.

These gain measurements are collected and binned in to HEALPix pixels,
creating an unpolarized antenna gain map for each horn pair. 

\subsection{Physical Optics Modeling}\label{section:dadra}

In addition to measurements, antenna gain patterns were modeled for
each feed horn position.  A physical optics code called Diffraction
Analysis of a Dual Reflector Antenna (DADRA), produced by YRS
associates \citep{YRS:DADRA}, calculates the full antenna gain and
polarization patterns from the profile of a corrugated feed horn, and
the precise positions and shapes of the primary and secondary reflectors.
Distortions of the reflectors away from their design shapes, measured on
the ground, are included in the calculation, as are the radiator
panels (modeled as flat, perfect conductors) behind the primary
reflector.  This calculation is discussed in more detail in
\citep{page/etal:2003,barnes/etal:2002}.

Physical optics calculations produce absolutely calibrated gains at a
single frequency.  Near the beam peaks, DADRA-predicted antenna
patterns agree within $\sim 3\%$ with gain measurements made in the
GEMAC test range.

Further from the main lobe, however, the DADRA predictions of the
sidelobe patterns are less accurate.  Bright sidelobes are predicted
in the right places, with the right shapes, but with lobe gains
incorrect by a varying factor of $\sim $ 0.5 to 2, compared to the
measurements.  As the ratio of predicted to measured gain varies across
a single lobe, this is not a calibration issue.  We find Princeton,
GEMAC and Lunar measured sidelobes in agreement where they overlap,
and in disagreement with the physical optics predictions.
 
The limitation is that DADRA's physical model of the spacecraft is too
simple.  The code cannot account for complicated self-reflecting
surfaces such as the front of the focal plane assembly.  Similarly, the
shape of the exposed aluminized Kapton shield around the secondary is
complex and not well known.  Neither of these surfaces is included in
the model.  For purposes of working with sidelobe gains, the physical
optics results must be considered a reasonable template from an
optical system similar to WMAP, rather than a precise model.

Nonetheless, the DADRA predictions prove extremely useful.  They yield
polarization orientations, difficult to extract from the measurements,
and provide evidence of systematic trends in the sidelobes.  For
example, the far sidelobe intensity in K-band is predicted to vary by
a factor of two across the band (from 18--25 GHz, with the longest
wavelengths giving the brightest lobes).  The DADRA code shows,
though, that the sidelobes maintain the same shape as frequency
varies: $G(\nu,\theta,\phi) \simeq N(\nu)g(\theta,\phi)$.  This is due
to the low edge-taper optical design.  This relation indicates that
single-frequency sidelobe measurements may be used to characterize
sidelobes, but their overall calibrations, even when available, may
not be assumed correct for calculating Galactic pickup.

Provided the main beams are accurate, this physical optics model may
also be used to generate broadband calibrations for the sidelobes.
Since 
\be
\int G(\nu,{\bf n}) \dee \Omega_{\bf n} = 4 \pi
\ee
for any lossless gain pattern, where $\nu$ is frequency and {\bf n} is
a direction on the sky, then one can write
\be
\int G(\nu,{\bf n})f(\nu) \dee \Omega_{\bf n}\dee\nu = 4 \pi \int f(\nu)\dee\nu \
,
\ee 
where $f(\nu)=r(\nu)T_{\rm sky}(\nu)$ carries all source and
radiometer frequency dependence.  Provided these are known, an
accurate physical optics model of the main beam allows for calibration
of the total power in the sidelobes:
\ba
\int_{\rm S}G(\nu,{\bf n})f(\nu) \dee \Omega_{\bf n}\dee\nu 
= 4 \pi \int f(\nu)\dee\nu   -  \nonumber \\
\int_{\rm M}G^{\rm model}(\nu,{\bf n})f(\nu) \dee \Omega_{\bf n}\dee\nu \ .
\labeq{calibrator}
\ea
Here M and S denote integration over the main beam and sidelobes, 
respectively.
If the main beam gain and source frequency spectrum are known,
equation~\eqn{calibrator} provides a means for calibrating 
measured sidelobe pickup.  The limitations of the physical optics
model do not materially affect main beam predictions, and so
the right hand side of equation~\eqn{calibrator} may be used to normalize
$G(\nu,{\bf n})$ on the left.  This calibration technique has been 
used for the K through V-band sidelobes.  In W-band, direct GEMAC 
measurements of power outside the main beam disagree with the modeled
value, suggesting that our model of the primary reflector does not 
correctly include its smallest scale distortions.  

\subsection{Combining Gain Measurements into an Overall Map}

\begin{figure*}
\epsscale{0.7}
\plotone{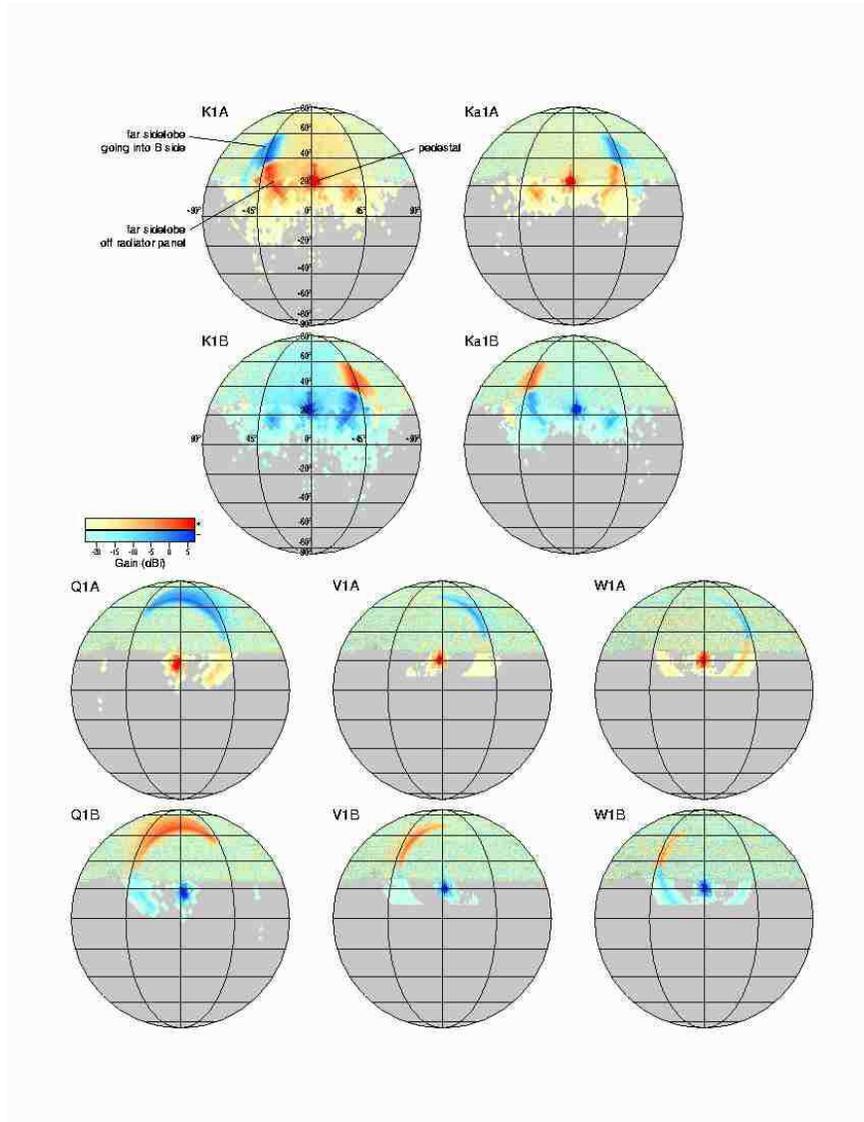}
\caption{Sidelobe intensity maps for each band. Two hemispheres are
shown in spacecraft coordinates for each differential antenna pattern.
Positive power (A-side) is shaded pink, negative pickup (B-side) is
shaded blue. Gray indicates regions where $|G| < $ -30 dBi. The main
beams lie at the center of each pedestal region (middle of the images,
at elevations of $\sim 20^\circ$).  The main beam gain is far off the
color scales, at +45 to +60 dBi for K through W bands.  The prominent rings of negative gain
on the A-side hemispheres come from radiation paths which miss the
edge of the B-side primary reflector, and reversely for the B-side
image. The V and W band maps were truncated at 10\deg of elevation
because the DADRA model is inaccurate below there.
\textbf{\textsl{
A higher quality rendering is available on the LAMBDA web
site.}}}
\label{fig:sidelobes}
\end{figure*}

To calculate sidelobe-induced Galactic pickup, a full $4 \pi$ sr
map of antenna gain and polarization is needed for each pair of
feed horns.  Such a map is most important for the antenna's highest
stray gain regions, all in the satellite's upper hemisphere.
Combined maps are constructed from the measurements
and models paying close attention to the
weaknesses of each data set, and the self-consistency of assembled
whole.  Several combined sidelobe antenna patterns are shown in Figure
\fig{sidelobes}.  The differing angular resolutions and noise levels 
are attributes of the different data sets. The Lunar data is prominent in the upper
third of the spheres.  The wide crescents of sensitivity, prominent
especially in Q-band, result from radiation spilling past the edge of the
primary reflector.

Full sky gain and polarization maps were assembled according 
to the following rules:

\smallskip  
In the immediate neighborhood of the beam peak (within a 
10\dg$\times$10\dg square centered on the beam peak), the GEMAC
measurements were used.  This region includes essentially all light
which reflects from the primary reflector.  The area is particularly
important, since it includes the ``beam pedestal,'' the area
immediately surrounding the main beam, produced by scattering off 
of the instrument.  Outside of the main beam, the pedestal is the 
highest gain region.

Where available, we use in-flight Lunar data.
Lunar measurements are the best available measurements of sidelobe
response.

Except at W-band, the overall sidelobe calibration is generated
from main beam physical optics predictions, via equation~\eqn{calibrator}.
W-band sidelobe maps were calibrated by assuming $T_{\rm Moon} = 175$
K, \citep{bennett/etal:1992a}.  This is lower bound on
the brightness for the range of lunar phases observed, and thus the 
W-band sidelobes are conservative.

Princeton gain measurements are used in regions away from
the beam peak, and where Lunar data is unavailable.
\smallskip

Relative calibrations are established by comparing high signal regions
where measurements overlap.  In K-band, the roughly
triangular high response region in Figure~\ref{fig:sidelobes} finds
the Lunar gain measurements are uniformly $60\%$ brighter than what
Princeton sees.  The Princeton and GEMAC measurements, both measured
at single nearby frequencies, are in agreement to within 10\%.
To generate a consistently
calibrated map, GEMAC and Princeton-measured gains were scaled up by
$30\%$, Lunar measurements scaled down by $30\%$.  This yields a
best-guess map for the whole sky with an overall calibration
uncertainty of $30\%$.  Similar considerations are necessary in each
band.  As our modeling of the beam and Moon observations mature, this 
uncertainty will be reduced.

Physical optics predictions are used in any region of
the sky not covered by the preceding measurements.  Except in V and
W bands, DADRA predictions are used exclusively in regions of the 
spacecraft's lower hemisphere containing no noticeable features.
Where the physical optics predictions are used, their
inaccuracy is harmless: in the lower hemisphere the ambient
predicted gains range from $-40$ to $-80$ dBi, K to W band.  On the
observatory, the body of the spacecraft (not present in the optical
model) tends to block out light incoming from below the spacecraft
horizons, so these gains are systematically high.  (The
gains are so low that varying them by a factor of 10
or even 100 has no measurable effect on sidelobe pickup.)  The
important lobes in the upper hemisphere are all at levels of 0 dBi or
slightly higher.  All measurements made with the source substantially
below the satellite horizon are limited by the measurement noise floor.
Preflight verification of the solar shield indicated that pickup the Sun's
position is rejected by greater than 90 dB.

Where only single-side measurements are available, the
differential gain pattern is constructed by reflection and rotation
of the single-side measurements.  That is, the differential gain 
is defined:
\ba 
G_{\rm diff,X}({\bf n}) & = & G_{\rm A,X}({\bf n}) - G_{B,X}({\bf n}) 
\nonumber\\
                       & = & G_{\rm A,X}({\bf n}) - G_{A,X}({\bf n}')\ ,
\ea
$$
\rm{where}\quad\quad {\bf n}'  =  \left(
\begin{array}{rrc}
 1  &   0  &  \ 0 \\
0   &  -1  &  \ 0 \\
0   &   0  &  \ 1 
\end{array}
\right) {\bf n} 
$$	
is a reflection in $y$, matching the A-B symmetry of the satellite.

Polarization directions are taken from the polarized optics
predictions, except in the immediate neighborhood of the beam peak.
In the pedestal region, high signal GEMAC measurements are used.  Two
measurements of pickup intensity are sufficient to determine the
antenna pattern polarization up to a sign ambiguity.  This ambiguity
is resolved by choosing the polarization direction closest to the one
predicted by DADRA.

Physical optics models indicate that the antenna pattern is
essentially linearly polarized at every point on the sky.  The
polarization ellipticity induced by the optics is negligible, both in
the main beam and sidelobes.  This result is unsurprising; it is
inevitable if one family of reflection/diffraction paths to the
telescope dominates from each point on the sky.  Only separate paths
of comparable strength can generate a circularly polarized component
to the beam.  To an excellent approximation the polarized
antenna pattern can be characterized by two fields:
$$
G_X({\bf n}),\ \mb{P}_X({\bf n})\ ,
\nonumber
$$
where $G$ is the gain, and $\mb{P}$ is a unit vector everywhere
perpendicular to the direction on the sky, ${\bf n}$. $\mb{P}$ carries
an unimportant sign ambiguity and uses three numbers to express a
single degree of freedom, but is coordinate independent and simplifies
calculation.

Although physical optics predictions get bright sidelobe
intensities only roughly correct, the generated polarization
directions should be almost exact, within $\sim$ 1\dg.  Since all the
bright sidelobes come from a single, clear reflection path --- e.g.,
horn to secondary, missing the primary and radiator panels --- even
geometric optics is sufficient to extract the polarization direction.
Differences between the physical optics predictions and the measured
antenna gains result from paths that are partially shadowed. This
shadowing does not significantly affect the polarization of the
remaining light.

\section{Calculating Sidelobe Pickup in Sky Maps}\label{section:calculation}

To calculate the sidelobe pickup, the 
antenna pattern is divided into main beam and sidelobe sections,
which are handled separately.  We choose to define the main beam to be
a circular region centered on the peak gain direction with a
cutoff radius, $\theta_{R_c}$, determined by Ruze theory predictions of
the scattering from the known distortions on the primary
\citep{page/etal:2003b}.  By band, these radii are K: 2.8\dg, Ka:
2.5\dg, Q: 2.2\dg, V: 1.8\dg, W: 1.5\dg.  The response within
$\theta_{R_c}$ is considered part of the main beam.   This response is
accounted for in the
map-making algorithms and included in the window functions.
Everything outside of $\theta_{R_c}$ is considered sidelobe gain,
treated here as a source of systematic offsets for temperature
measurements.  The ``beam pedestals'' are outside of $\theta_{R_c}$, and are
included in the sidelobes.
 
When the spacecraft is in an orientation ${\cal O}$ with respect to the
Galaxy, where ${\cal O}$ is the rotation matrix from Galactic
coordinates into the spacecraft
frame, the sidelobe pickup for differencing
assembly (DA) $X$ is:
\ba
\delta T_X({\cal O}) = {1\over 4\pi}\int r_{{}_X}(\nu)G_X({\cal O}{\bf n},\nu) \biggl[ 
T_{\rm sky}({\bf n},\nu)
+ {} \nonumber \\
 T_{\rm pol}({\bf n},\nu)
\left\{\mb{P}_{\rm sky}({\bf n})\cdot {\cal O}^\top \mb{P}_X({\cal O}{\bf n})\right\}^2
\biggr]\ \dee \Omega_{\bf n}\, \dee\nu\ . 
\labeq{systematic}
\ea
Here $T_{\rm sky}(\nu)$ is the effective unpolarized sky temperature,
$T_{\rm pol},\mb{P}_{\rm sky}$ are a temperature and unit vector
characterizing the linearly polarized portion of the sky, and $G_X$
has been set to zero within $\theta_{R_c}$.  Here $r_{{}_X}(\nu)$ is the
radiometer gain, normalized so that $\int \dee\nu\, r_{{}_X} =1$.  For
calculations throughout the paper, we will use uncorrected map
temperatures $T_{\rm map}$ as an estimate for $T_{\rm sky}$.

In order to calculate $\delta T$ in equation~\eqn{systematic}, 
we make the following approximation:
Within a band, we integrate out the 
frequency dependence of the sky,
\ba
\int {\rm d \nu}\, r_{{}_X}(\nu)G_X({\bf n},\nu)T_{\rm sky}({\bf n}',\nu)
= {}\nonumber \\
\quad \quad g_X({\bf n})T_{\rm sky,\mit X}({\bf n}')\ ,\quad \rm and
\labeq{unpolfreqint}
\ea
\ba
\int {\rm d \nu}\, r_{{}_X}(\nu)G_X({\bf n},\nu)T_{\rm pol}({\bf n}',\nu)
= {} \nonumber \\
\quad\quad g_X({\bf n})T_{\rm pol,\mit X}({\bf n}')\ ,
\labeq{polfreqint}
\ea
for all directions ${\bf n},{\bf n}'$.  This is a reasonable
approximation, since the microwave sky maps $T_{\rm sky,\mit X}$ were
constructed using the same differential radiometers which see the
sidelobes.  The frequency variation of the sidelobe pattern itself,
$G_X({\bf n},\nu) \simeq N(\nu)g_X({\bf n})$ amounts to an overall
calibration uncertainty in the sidelobe strength.  This is correct
provided that the spectrum of the source of stray light pickup is
uniform across the sky. Of course, the various foreground spectra
\citep{bennett/etal:2003c} are manifestly {\it not} uniform around the
sky as frequency ranges from K to W-bands, 22 to 100 GHz. However,
within any particular band (e.g., K band) the brightest foregrounds
tend to be dominated by components with similar spectra (e.g.,
synchrotron radiation.) It is likely that the spectrum of the
brightest polarized foregrounds in a band differs from their
unpolarized counterparts, so $g_X$ may be differently calibrated in
equations~\eqn{unpolfreqint} and \eqn{polfreqint}.

The relevant question for \WMAP is: How does the differential sidelobe
pickup $\delta T_X({\cal O})$ in equation~\eqn{systematic} contribute to 
microwave sky maps?  Each measurement in the data stream contributes
to its sky map at exactly two pixels: the A and B-side main beam
line of sight directions.  For a given pixel $p$, the sidelobe contamination 
may be calculated by taking the appropriate average of $\delta
T_X({\cal O}_p)$ for all spacecraft orientations ${\cal O}_p$ where
the A or B-side beams land on $p$.  This average over spacecraft
orientations must be weighted according to the flight scan pattern,
and should reproduce the result of the full iterative map-making
algorithm as closely as possible.  Maps of unpolarized and polarized 
contamination may be generated from the same basic approach.

\subsection{Unpolarized Sidelobe Pickup}\label{section:unpolarized}

The radiometer data separates cleanly into polarized and unpolarized
components, so the two terms in equation~\eqn{systematic} may be
handled separately.  For unpolarized pickup, ($T_{\rm pol}=0$):
\be
\delta T_X({\cal O}) = {1 \over 4\pi} \int g_X({\cal O}{\bf n})T_{\rm sky,\mit
X}({\bf n})\, \dee\Omega_{\bf n}\ .
\labeq{unpol_pickup}
\ee
Since the sidelobes are broad and smooth, the integral in
equation~\eqn{unpol_pickup} is calculated accurately using a relatively
coarse pixelization of the sphere, HEALPix $N_{\rm
side}={}$ 32 or 64.  From these varying sidelobe signal in the differential
measurements, one can extract the induced systematic contamination of the 
microwave sky map.  We use two techniques to approximate pickup in the
final map.  Both follow the mapmaking algorithm from the flight data stream.

Applying the first iteration of the mapmaking
algorithm \citep{hinshaw/etal:2003b}, one can write:
\ba
\delta T_{\rm ind\mit,X}({\bf n}) & \simeq & {1\over 2}\left\langle\delta
T_X({\cal O})\right\rangle_{{\cal O}:{\cal O}{\bf A}_X={\bf n}} \nonumber \\
& & -{1\over 2}\left\langle\delta T_X({\cal O})\right\rangle_{{\cal O}:{\cal O}{\bf B}_X={\bf n}}\ ,
\labeq{chris_approx}
\ea
where $\left\langle\right\rangle_{\cal O}$ denotes the average over
the one-dimensional family of all spacecraft orientations ${\cal O}$
with a main beam pointing at pixel ${\bf n}$. The average
$\left\langle\right\rangle_{\cal O}$ is weighted according to the
distribution of in-flight spacecraft orientations. For example, pixels
near the ecliptic poles are sampled almost evenly among possible
spacecraft orientations.  For pixels near the ecliptic equator, the
flight scan pattern restricts the spacecraft orientation into two disjoint
sets, and the full range of orientations is never observed.
Similarly, in the full analysis the time ordered data is blanked whenever
a main beam crosses a planet, or the reference beam crosses the Galaxy
or a bright point source.   Spacecraft orientations here are weighted
using exactly the same cuts as are used in the full map-making analysis.

Equation~\eqn{chris_approx} yields close to the correct sidelobe
contribution, since most of the power in each sky pixel is generated
by the first iteration of the mapmaking algorithm
\citep{hinshaw/etal:2003b}.  However, it is imperfect, since
iterations $2$ and higher do have some effect on the data.  In
particular, the pickup from the bright reference beam pedestal (the B
pedestal, if one is averaging over constant A beam peak direction)
over-contributes to equation~\eqn{chris_approx}.  Multiple iterations
of the map-making algorithm separate out any power symmetrically
localized around the reference pixel, removing most of the reference
beam pedestal pickup from the map.

The second method of calculating the map contamination skips the
reverse-differencing algorithm altogether.  Separating the A and B
side pickup according to sign:
\ba
 g_{{}_{A,X}}({\bf n}) &=& \left\{ \begin{array}{rcl} 
g_X({\bf n}) &:&    g_X({\bf n}) > 0 \\
0           &:&     g_X({\bf n}) < 0 
 \end{array} \right .
\nonumber \\                   
g_{{}_{B,X}}({\bf n}) &=& \left\{ \begin{array}{rcl} 
0           &:&    g_X({\bf n}) > 0 \\
-g_X({\bf n}) &:&     g_X({\bf n}) < 0 
 \end{array} \right .
\nonumber                    
\ea
one can then write a non-differential mean sidelobe pickup:
\ba
\delta T_{\rm ind,X}({\bf n})  \simeq  \hbox{\hskip 1.in}\nonumber \\
 {1\over 8\pi} 
\left\langle\int T_{\rm sky,X}({\bf n}')g_{{}_{A,X}}({\cal O}{\bf n}') {\rm d
\Omega_{{\bf n}'}}\right\rangle_{{\cal O}:{\cal O}{\bf A}_X={\bf n}} + {}\nonumber \\
 {1\over 8\pi} 
\left\langle\int T_{\rm sky,X}({\bf n}')g_{{}_{B,X}}({\cal O}{\bf n}') {\rm d
\Omega_{{\bf n}'}}\right\rangle_{{\cal O}:{\cal O}{\bf B}_X={\bf n}}\ .
\labeq{gary_approx}
\ea
No reference beam pedestal ever shows up in
equation~\eqn{gary_approx}, and so the largest difficulty in the
approximation of equation~\eqn{chris_approx} is avoided. This
approximation would be exactly right if the instrument yielded the sum
of two independent telescope measurements, A and B.  If the antenna
patterns $g_{{}_{A,X}},g_{{}_{B,X}}$ were symmetric about the A and B
side beam peaks, the complete map-making algorithm would give exactly
such a sum.  The idea of equation~\eqn{gary_approx} 
is to jump immediately to the end result of
map-making.

It is evident from Figure~\fig{sidelobes} that the antenna gain
patterns are not axially symmetric around their beam peaks.  The beam
pedestals are somewhat asymmetric, and the far sidelobes show no axial
symmetry at all.  Since equation~\eqn{gary_approx} removes any regions
of negative gain from the beam pattern, cancellations between negative
and positive pickup which lower the differential signals do not occur,
and the resulting $\delta T$ is biased high. For the same reason,
equation~\eqn{gary_approx} is problematic when applied to noisy differential
data (the Lunar signal.)  Separating differential gain according
to sign introduces a bias which again overestimates the sidelobe signal.

Nevertheless, for the \WMAP\ data both sidelobe estimation techniques
produce reasonable results.  They agree to within a few tens of
percent across the sky, with equation~ \eqn{chris_approx} showing more
variation due to reference beam pedestal pickup.
Equation~\eqn{gary_approx} would be correct for beam patterns which
are axially symmetric about the line of sight, and
equation~\eqn{chris_approx} maximally includes pickup from axial
asymmetry.  The true pattern can be written as a weighted sum of
symmetric and asymmetric pieces.  Since the symmetric and asymmetric
weights are comparable, the best estimate sidelobe contribution was
chosen to be the mean of equations~\eqn{chris_approx} and
\eqn{gary_approx}.  This mean will be closer to the true pickup than
either approximation separately.

\subsubsection{Results for Unpolarized Sidelobe Pickup}

\begin{figure*}
\epsscale{1.0}
\plotone{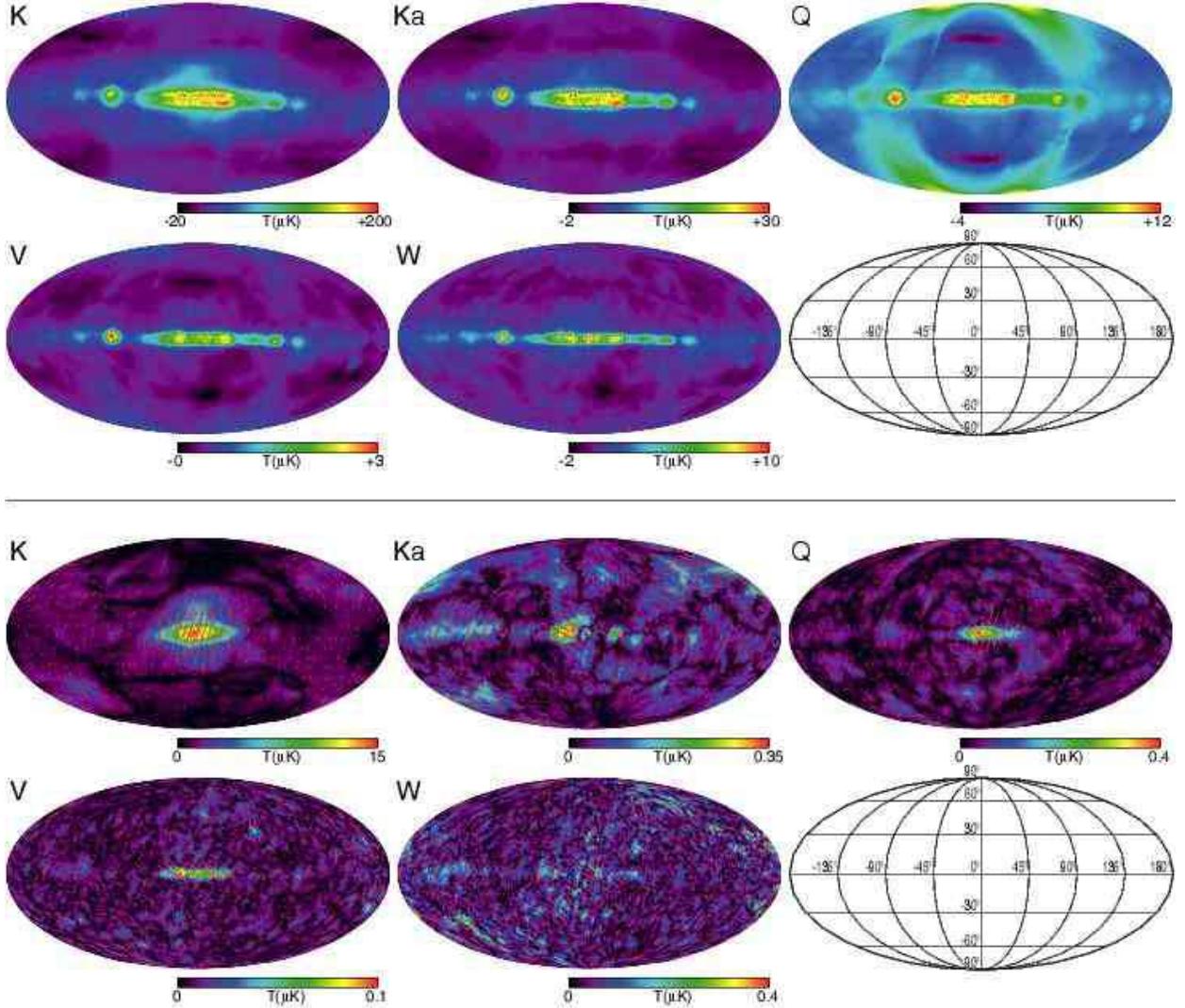}
\caption{ Top: Unpolarized sidelobe pickup contamination for selected 
sky maps. For purposes of CMB analysis, sidelobe contributions far
from the Galactic plane are most important. Bottom: Polarized sidelobe
contamination of the CMB maps.
  Intensity $(Q^2+U^2)^{1/2}$ is shown as a color scale, while polarization
directions appear as red lines.  
{\em These are sidelobe contamination maps,  and should not be used to
extract information about the polarized microwave sky.  They are
dominated by signal from the radiometer bandpass mismatch.}
\textbf{\textsl{
A higher quality rendering is available on the LAMBDA web site.}
}}
\label{fig:allmaps}
\end{figure*}

Figure~\fig{allmaps} shows the unpolarized sidelobe contributions to 
sky maps for K through W-band.  As expected, these images are
dominated by large-scale power, and are everywhere much weaker than
the CMB signal they contaminate.  K-band, with the strongest
sidelobes looking at the brightest Galactic foregrounds, has
the most sidelobe pickup.
Detailed averages for each differencing assembly are
listed in Table~\ref{table:unpolarized_results}.

\begin{table*}
\begin{center}
\begin{tabular}{lcccccc}
\hline\hline
DA &  Mean & Min &  Max & rms & $\ell_{\rm max}$ & $\max(C_\ell)$\\
   & ($\mu$K) &  ($\mu$K) &  ($\mu$K) &  ($\mu$K) & &  $(\mu$K$^2)$ \\ 
\\
\hline
K1 &     9   &   -17   &    72   &    15   & 6 &   30    \\ 
Ka1 &     2   &   -1.6   &     9   &     2   & 6 &   0.4    \\ 
Q1 &    1.4   &    -4   &    10   &     2   & 2 &    4    \\ 
Q2 &    1.3   &    -4   &    10   &     2   & 2 &    4    \\ 
V1 &    0.3   &   $ -2 \times 10^{-2}$   &    0.6   &    0.3   & 2 &  $  3 \times 10^{-2}$    \\ 
V2 &    0.2   &   $ -2 \times 10^{-2}$   &    0.6   &    0.2   & 2 &   $  2 \times 10^{-2}$    \\ 
W1 &   -0.12   &   -1.4   &    1.0   &    0.4   & 4 &  $  6 \times 10^{-2}$    \\ 
W2 &   $ -6 \times 10^{-2}$   &    -3   &     3   &    0.8   & 4 &   0.5    \\ 
W3 &   $ -5 \times 10^{-2}$   &    -3   &     3   &    0.8   & 4 &   0.5    \\ 
W4 &   -0.12   &   -1.4   &    1.0   &    0.4   & 4 &  $  8 \times 10^{-2}$    \\ 
\hline
\end{tabular}
\end{center}
\label{table:unpolarized_results}
\caption{Sidelobe contamination levels for unpolarized microwave
sky maps.  
These averages were taken in the CMB analysis region, specifically
outside of the Kp0 Galaxy + source mask. Here $C_\ell$ are angular
power for the sidelobe pickup maps autocorrelation.
}
\end{table*}

Figure~\fig{unpol_spectrum} shows angular power spectra for the sidelobe
pickup contamination for each DA.  Each spectrum is dominated
by its lowest $\ell$ components, $0 \le \ell \lsim 10$.  With the
exception of K-band, the spectra lie below $\sim 5 (\mu\rm K)^2$.
Since the true sky signal dominates over the sidelobe contamination in
every band, the cross-correlation between 
the sidelobe-pickup maps and the CMB is more relevant.  The $C_\ell$ are
extracted from 
\be 
T_{\rm map} = T_{\rm sky} + T_{\rm contam} + {\rm noise}\ ,
\ee
so for a small contamination signal, the leading effect in the power
spectrum comes from the $T_{\rm sky}\times T_{\rm contam}$ term.
These cross-correlations are shown in Figure~\fig{unpol_crosscorr}.
K-band shows a contribution around $100\ (\muK)^2$ at low $\ell$, while
all other bands are below $20\ (\muK)^2$.

\subsection{Polarized Sidelobe Pickup}\label{section:polarized}

Extracting polarized pickup due to sidelobes is similar to the
unpolarized case, although the process is more involved. There are two
channels of differential data per pair of feed horns, each one
differencing one polarization from the A-side horn with the opposite
polarization from B-side.  That is, the signals are: $X1 = (A_- -
B_|)$ and $X2 = (A_|-B_-)$, where $A_{|(-)}$ is power entering the
radiometer from the $|$ $(-)$ polarized arm of the OMT on the A-side
feed horn, and similarly for $B_{-(|)}$.  The unpolarized signal
channel is $X1+X2$, while the purely polarized signal is
\be
X1-X2 = (A_- - A_|) - (B_| -B_-)\ .
\labeq{pol_signal}
\ee
If the antenna patterns of the two perpendicular polarizations seen
through the same feed horn were different on the sky, unpolarized
pickup of a varying sky could be misidentified as a polarized
signal.  Fortunately, 
the sidelobe gain patterns closely match for the two opposite
polarizations, in both measurements and models.
That is
\ba
g_{X-}({\bf n}) &=& g_{X|}({\bf n})\ , \quad {\rm and} \\
\mb{P}_{X-}({\bf n}) &=& {\bf n} \times \mb{P}_{X|} \ ,
\labeq{match}
\ea
for all ${\bf n}$, to within the uncertainties of the models and 
measurements.
A direct test of this match was to
take a polarized source, measure the gain pattern through
one arm of a feed horn OMT, rotated the source by 90\deg and 
switch the receiver to the other arm of the OMT, and repeated the
measurement. This always generated the same antenna pattern up to measurement
uncertainty.
Small main beam deviations from
equation~\eqn{match} were observed; these are reported and discussed in 
\cite{page/etal:2003b}.

There is, however, a mechanism by which {\it unpolarized} foreground
pickup can induce a polarized signal, equation~\eqn{pol_signal}.  The mismatch
comes not in the optics, but in the broadband radiometer gains,
$r_{{}_{X1}}(\nu),\ r_{{}_{X2}}(\nu)$.  The radiometers are separately
calibrated on the CMB dipole, and so will agree on any CMB signal.
However, if a source has a different spectrum,
for example a steeply non-thermal $T_{\rm sky}({\bf n},\nu) \propto \nu^{\alpha-2}$
\citep{page/etal:2003b}, with $\alpha-2$ significantly far from zero, then
any mismatch in radiometer gains can generate a spurious polarization
signal:
\be
\delta T_{{\rm pol},X}^{\rm spr} = \int \dee \nu\,
\left(r_{{}_{X1}}(\nu)-r_{{}_{X2}}(\nu)\right)T_{\rm sky}^{\rm nt}(\nu)
\labeq{spurious_pol}\ .
\ee
Here $T_{\rm sky}^{\rm nc}$ is any non-CMB sky signal, and
spacecraft orientation and the integral over the sky have been
suppressed.  The signal in equation~\eqn{spurious_pol} comes from inherently
unpolarized foreground sources, and so pickup from it is independent
of the spacecraft polarization direction, i.e., there is no
$\mb{P}_{\rm sky}\cdot {\cal O}^\top \mb{P}_X \sim \cos(2\Phi_{\rm sc,sky})$
dependence in equation~\eqn{spurious_pol}.  In the limit of even
sampling over all spacecraft orientations, main beam pickup of this
uniform signal would not contribute to the polarization maps.
However, the non-uniformity of the actual set of spacecraft
orientations does cause this main-beam spurious pickup to leak into the
polarization maps; see \cite{kogut/etal:2003} for a full discussion.

Through sidelobe pickup, however, this spurious polarization signal 
directly contributes to the polarized maps, even where spacecraft
orientation is sampled uniformly. When the A-side main beam
points at a particular pixel, the sidelobe pickup from equation~\eqn{spurious_pol}
will naturally vary with spacecraft orientation $\Phi$, since the 
sidelobes will swing across the foregrounds as the spacecraft
is rotated about the main-beam axis.  This spurious pickup must be
included along with the pickup from polarized sources to calculate the
sidelobe-induced foreground contamination of our polarization maps.
Fortunately it is possible to accurately calculate this pickup for
each radiometer.  If one takes the polarized radiometer signal, 
equation~\eqn{pol_signal}, from the time ordered data, and then runs it through the 
{\it unpolarized} map-making algorithm, one obtains a full sky map:
\be
T_{{\rm pol},X}^{\rm spr}({\bf n}) = \left[\mbox{Map-making}\right]\circ
(X1-X2)_{\rm TOD}\ ,
\labeq{spurious_sky}
\ee
which is composed purely of the spurious polarization signal for the
radiometer pair $X1,X2$.  Genuine polarized pickup enters the
time ordered data of $(X1-X2)$ equally with opposite signs, and so is null in
of equation~\eqn{spurious_sky}.  The sidelobe pattern can then
be folded back in with this sky map for each radiometer pair to 
calculate the spurious signal for any spacecraft orientation.

\smallskip

Polarized sidelobe pickup results from two sources: polarized sky
signal in polarized sidelobes, and radiometer RF band mismatch on
foregrounds.  The most significant pickup comes
from the bright plane of the Galaxy and the band mismatch, followed by
the strongly polarized foregrounds in the plane, a
couple of supernova remnants, and the Northern Galactic Spur.

To generate maps of sidelobe-induced contamination in the sky maps,
again we start with the existing year one polarized sky maps, for each
horn pair (a single differencing assembly).  When the telescope is in
orientation ${\cal O}$, the polarized sky signal it sees through the
sidelobes is:
\ba
\delta T_{\rm pol,\mit X}({\cal O}) 
= {1\over 4 \pi} 
\int \dee\Omega_{\bf n}\  g_X({\cal O}{\bf n})T_{\rm pol,\mit X}({\bf n})\times
\nonumber \\
\biggl[ \left(\mb{P}_{\rm sky,\mit X}({\bf n})\cdot {\cal O}^\top\mb{P}_{X-}({\cal O}{\bf n})\right)^2 - 
\nonumber \\
\left(\mb{P}_{\rm sky,\mit X}({\bf n})\cdot {\cal O}^\top\mb{P}_{X|}({\cal O}{\bf n})\right)^2
\biggr] \nonumber \\
+ {1\over 4 \pi} 
\int \dee\Omega_{\bf n}\  g_X({\cal O}{\bf n})T_{\rm pol,\mit X}^{\rm spr}({\bf n})
\ .
\ea
Since $\mb{P}_-$ is everywhere perpendicular to $\mb{P}_|$, the
integral may be written
\ba
\delta T_{\rm pol,\mit X}({\cal O}) = {1\over 4 \pi} 
\int \dee\Omega_{\bf n}\  g_X({\cal O}{\bf n}) \times  \nonumber \\ 
\left[T_{\rm pol,\mit X}({\bf n})\cos \left(2\Phi({\bf n},{\cal O})\right) 
+ T_{\rm pol,\mit X}^{\rm spr}({\bf n})
\right]
\ ,
\labeq{dtpol}
\ea
where $\Phi$ is the angle between incoming sky polarization and 
antenna pattern polarization, for a particular point on the sky and 
spacecraft orientation.

\newcommand \dtpol {\delta T_{\rm pol,\mit X}({\cal O})}
\newcommand \dtpolind {\delta T_{\rm pol,\mit X}^{\rm ind}({\bf n})}

Each measurement $\dtpol$ contributes to the recovered polarization of
two pixels, the positions of the A and B beam peaks at orientation
${\cal O}$.  For polarization, the angle, $\gamma$, between beam peak
polarization direction and the chosen set of sky coordinates is also
needed.  For \WMAP's maps of the polarized sky, we chose to reference
Stokes parameters to the Galactic meridian.  Here define
$\gamma_X({\cal O})$ as the angle from the instrument polarization
vector ${\cal O}^\top\mb{P}_X$ to the Galactic meridian.

Tracing the sidelobe pickup $\dtpol$ through the map-making algorithm,
the data pipeline receives a set of offsets to measurements of an
unknown polarization intensity and direction at a particular pixel,
${\bf n} = {\cal O}^\top\mb{A}_X$ or $ {\cal O}^\top\mb{B}_X$.  These values correspond
to a series of $n$ measurements of a local Stokes $Q$ parameter, in a
frame rotated by an angle $\gamma_{A(B),X}({\cal O})$ from the frame
where we wish to extract $Q$ and $U$.  That is, it sees a series of
$n$ measurements $Q_i$:
\be
\left[\begin{array}{c} Q_1 \\ Q_2 \\ \vdots \\ Q_n \end{array}\right]
= \left[ \begin{array}{cc} \cos 2\gamma_1 & -\sin 2\gamma_1 \\
                       \cos 2\gamma_2 & -\sin 2\gamma_2 \\
                               \vdots & \vdots \\
                       \cos 2\gamma_n & -\sin 2\gamma_n \end{array}\right]
\left[\begin{array}{c} Q \\ U \end{array} \right] \ ,
\ee
where $Q,U$ are the unknown, desired values.
The values of $Q,U$ which best fit this data set are:
\be
\left[\begin{array}{c} Q \\ U \end{array}\right] = M^{-1}
\sum_i \left[\begin{array}{r} 
 Q_i\cos 2\gamma_i \\ - Q_i\sin 2\gamma_i 
\end{array}\right]\ ,
\ee
where 
\be
M  =  \sum_i \left[\begin{array}{rr}
 \cos^2 2\gamma_i & - {1\over 2} \sin 4\gamma_i \\
 - {1\over 2} \sin 4\gamma_i &  \cos^2 2\gamma_i 
\end{array}\right]\ .
\labeq{Mdef}
\ee

This is a linear system, so the induced sidelobe pickup travels
through the mapmaking procedure independently from any main beam
signal.  After generating a family of $\dtpol$ by performing the
integrals in equation~\eqn{dtpol}, one can then construct the induced
contributions to the $Q$ and $U$ maps.
\be
\left[\begin{array}{c}
\delta Q^{\rm ind}_X({\bf n}) \\
\delta U^{\rm ind}_X({\bf n}) \\
\end{array}\right] \simeq {1\over 2}\left[\begin{array}{c}
\delta Q_{A,X}({\bf n}) +\delta Q_{B,X}({\bf n}) \\
\delta U_{A,X}({\bf n}) +\delta U_{B,X}({\bf n}) \\
\end{array}\right]\ ,
\labeq{polavg}
\ee
where 
\ba
\left[\begin{array}{c}
\delta Q_{A,X}({\bf n}) \\
\delta U_{A,X}({\bf n}) \\
\end{array}\right] =  {1\over 2} M^{-1}_{A,X}\times \hbox{\hskip 1.in}
\nonumber \\ 
\sum_i \left[\begin{array}{r} 
 \delta T_{\rm pol,\mit X}({\cal O}_i) \cos \left(2\gamma_{{}_{A,X}}({\cal O}_i,{\bf n})\right) \\ 
-\delta T_{\rm pol,\mit X}({\cal O}_i) \sin \left(2\gamma_{{}_{A,X}}({\cal O}_i,{\bf n})\right) 
\end{array}\right]\ ,
\labeq{poloffset}
\ea
and similarly for the B side, with the sign of $\dtpol$ reversed. Here
$M_{A,X}$ is defined as in equation~\eqn{Mdef}, for the set of angles:
$\left\{\gamma_{{}_{A,X}}({\cal O}_i,{\bf n})\right\}$.  All ${\cal
O}_i$ preserve the pixel whose polarization is under study: ${\cal
O}_i{\bf n} = {\bf n}$.  As for the unpolarized maps, it is crucial to
use the same family of ${\bf n}_i$ present in the actual observations
in the first year scan pattern.  The new factor of $1/2$ in
equation~\eqn{poloffset} comes from the map-making assumption that one
half of the measured signal comes from each side.

In the end, equation~\eqn{poloffset} approximates the polarized
sidelobe pickup with a method precisely analogous to
equation~\eqn{chris_approx} for overall power.  From the existing
microwave maps and antenna patterns, one creates a family of simulated
measurement differences $\delta T_{\rm pol}$, and then propagates them through one iteration of
the map-making algorithm.  As before, to propagate the simulated data
through the full map-making data pipeline is a task for the next
generation of data analysis.

\subsubsection{Results for Polarized Sidelobe Pickup}

The lower portion of Figure~\fig{allmaps} shows the sidelobe
contamination in one map for each band.  The ``spurious'' pickup
due to radiometer bandpass mismatch slightly dominates true polarized
pickup, and so the Galaxy, appearing as a coherent, polarized source, 
dominates the images.  Even the brightest sidelobe contamination is
extremely weak: the Galaxy peaks at $\sim 400$ nK in all bands except
K, where a stronger band mismatch leads to a 16\ \muK~signal in the plane.

\begin{table*}
\begin{center}
\begin{tabular}{lcccc}
\hline\hline
DA &  Mean & Min &  Max & rms \\
   & ($\mu$K) &  ($\mu$K) &  ($\mu$K) &  ($\mu$K) \\ 
\\
\hline
K1 &    0.8   &   $  7 \times 10^{-3}$   &     5   &    1.0   \\ 
Ka1 &   $  3 \times 10^{-2}$   &   $  3 \times 10^{-4}$   &    0.13   &   $  4 \times 10^{-2}$   \\ 
Q1 &   $ 1.6 \times 10^{-2}$   &   $  8 \times 10^{-5}$   &   $  9 \times 10^{-2}$   &   $  2 \times 10^{-2}$   \\ 
Q2 &   $  3 \times 10^{-2}$   &   $  9 \times 10^{-5}$   &    0.3   &   $  5 \times 10^{-2}$   \\ 
V1 &   $  2 \times 10^{-3}$   &   $  3 \times 10^{-5}$   &   $ 1.4 \times 10^{-2}$   &   $  3 \times 10^{-3}$   \\ 
V2 &   $  2 \times 10^{-3}$   &   $  3 \times 10^{-5}$   &   $ 1.4 \times 10^{-2}$   &   $  2 \times 10^{-3}$   \\ 
W1 &   $  6 \times 10^{-2}$   &   $  9 \times 10^{-5}$   &    0.5   &   $  7 \times 10^{-2}$   \\ 
W2 &   $  4 \times 10^{-2}$   &   $  4 \times 10^{-4}$   &    0.2   &   $  5 \times 10^{-2}$   \\ 
W3 &   $  4 \times 10^{-2}$   &   $  4 \times 10^{-4}$   &    0.2   &   $  5 \times 10^{-2}$   \\ 
W4 &   $  8 \times 10^{-2}$   &   $1.0 \times 10^{ -3}$   &    0.5   &   $  9 \times 10^{-2}$   \\ 
\hline
\end{tabular}
\end{center}
\caption{Contamination of the polarized maps due to sidelobe
pickup. Averages are taken outside of the Kp0 mask region, away from the
Galaxy. }
\label{table:polarized_results}
\end{table*}

Table~\ref{table:polarized_results} shows means for the sidelobe
contamination of each Q,U pair of polarized maps. With the exception
of K-band, where a stronger bandpass mismatch leads to a polarized
signal ranging to 5\ \muK\ outside of the Kp0 mask, expected
contamination per pixel is $\lsim 400 $ nK.  Angular power spectra for
the intensity of polarized sidelobe contamination are shown in
Figure~\fig{pol_spectrum}.

Again, the power spectrum of the sidelobe contamination itself is less
interesting than its cross-correlation with the CMB.  Polarized
sidelobe pickup contributes to the TE power spectrum, most strongly
via the term $T_{\rm sky}\times E_{\rm contam}$, where $E_{\rm
contam}$ is generated from the $Q$ and $U$ sidelobe contamination
maps.  Applying the same analysis as is used for the microwave sky TE
power spectrum \citep{kogut/etal:2003}, one can calculate the
sidelobe-induced contribution to the \WMAP year one TE spectra.  The
CMB $\times$ polarized sidelobe map TE spectra are shown in
Figure~\fig{TE_spectrum}.  In K-band, sidelobe polarization pickup
generates a TE signal comparable to our reported value, $\ell
|X_\ell|$ ranges from 1 to 10\ $\muK^2$.  For all other bands,, the
cross-correlation angular power $\ell|X_\ell|$ ranges from .03 to 0.5
~$\muK^2$, with a roughly flat spectrum.  K-band polarization maps are
corrected for sidelobe pickup before they are used to calculate CMB
spectra.  For all other bands, sidelobe contamination is a subdominant
contributor to the TE spectrum error.

\begin{figure*}
\begin{tabular}{p{3.0in}p{3.0in}}
\includegraphics[width=3.0in]{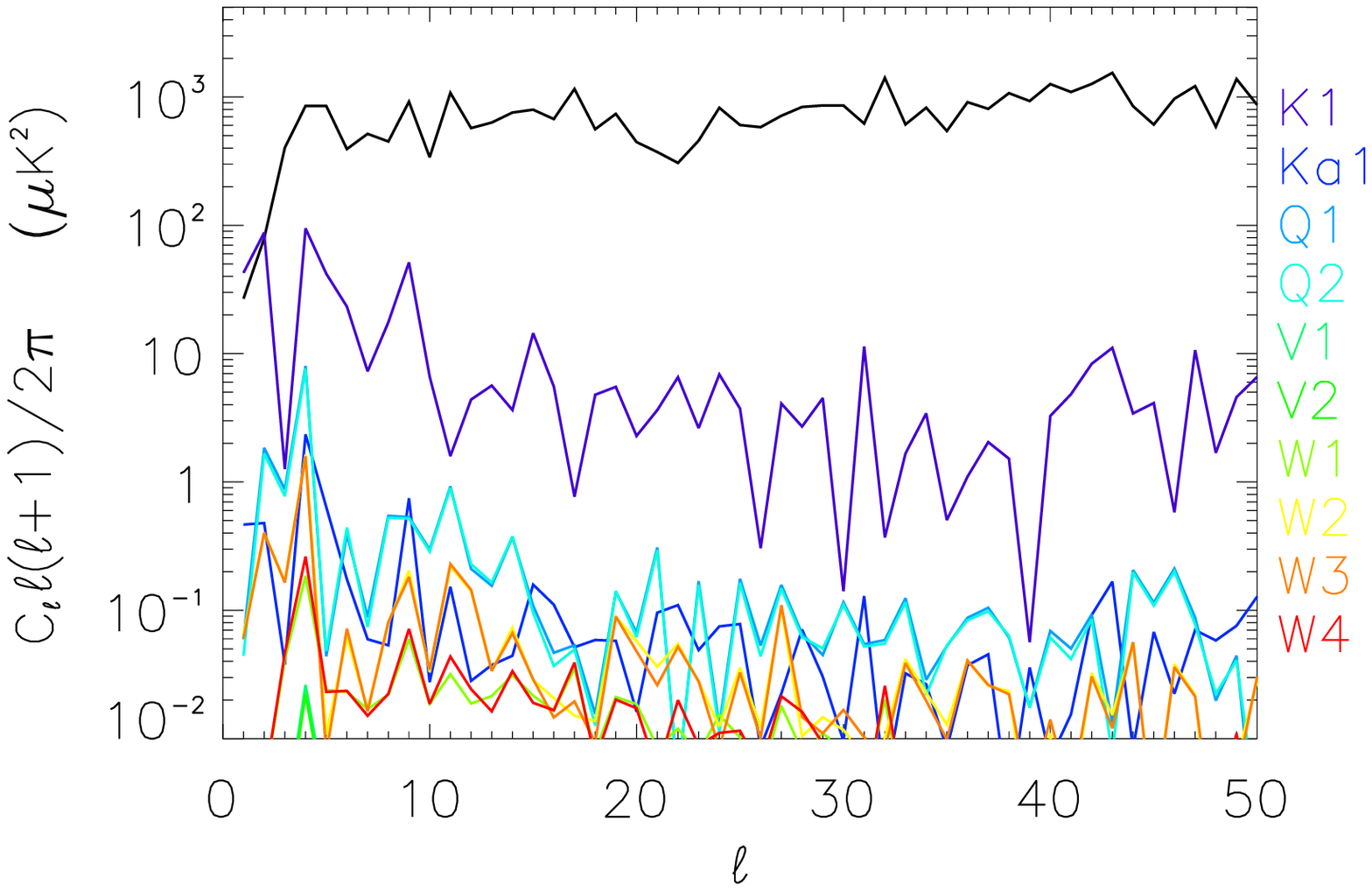}
&
\includegraphics[width=3.0in]{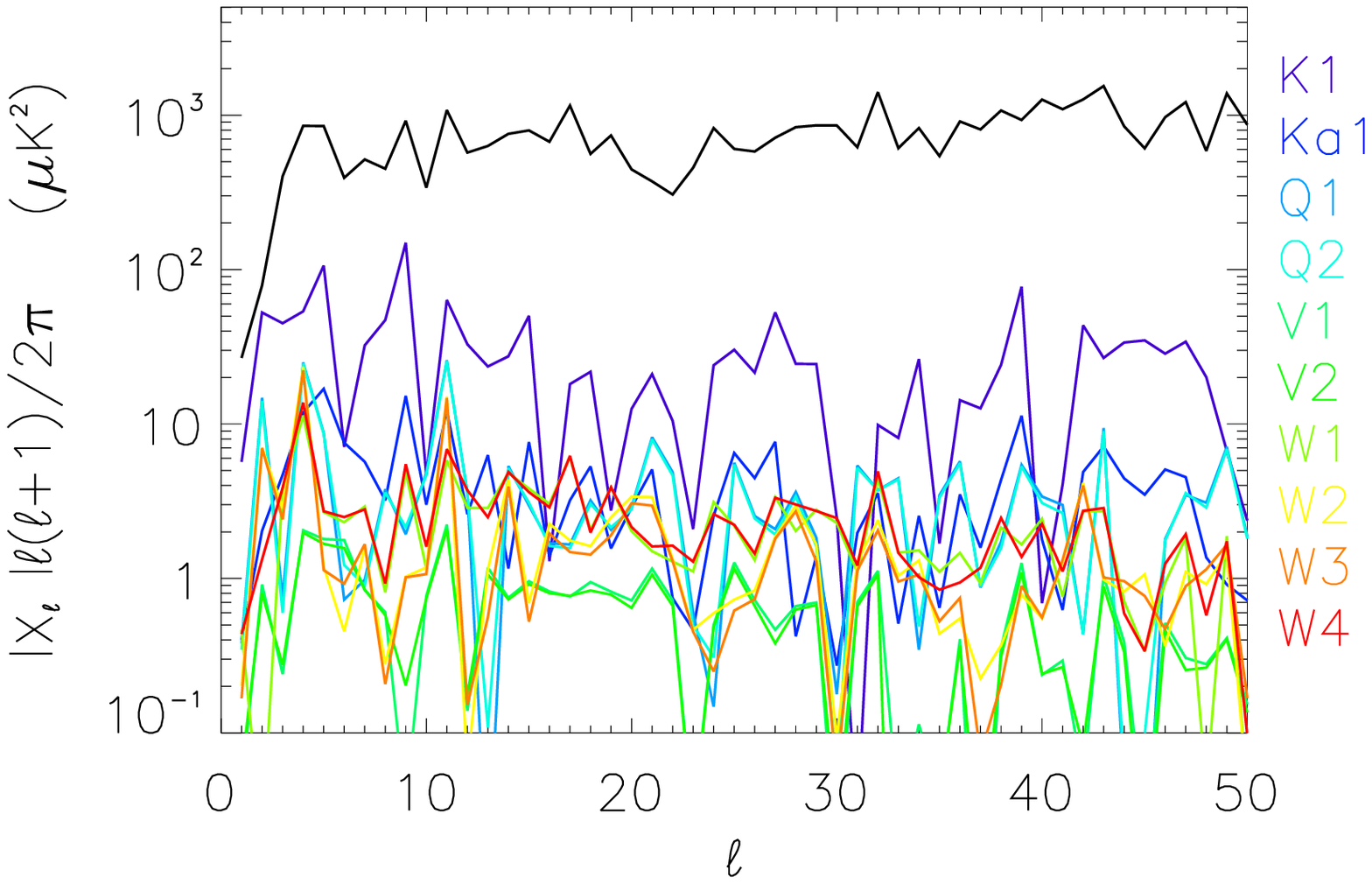}
\\
\caption{Angular power spectra of sidelobe contamination maps for the {\sl WMAP}  year one 
sky maps.  A CMB angular power spectrum is shown in black for comparison.
All spectra shown were made with the Kp0 Galaxy+source mask 
\protect{\citep{bennett/etal:2003b}}.}
\labfig{unpol_spectrum}
&
\caption{Absolute value of the cross-correlation between sidelobe 
pickup and the microwave sky.  This is the extent to which sidelobe 
contributions should contaminated calculated $C_\ell$ values.
A CMB power spectrum is shown in black for comparison.}
\labfig{unpol_crosscorr}
\\
\includegraphics[width=3.0in]{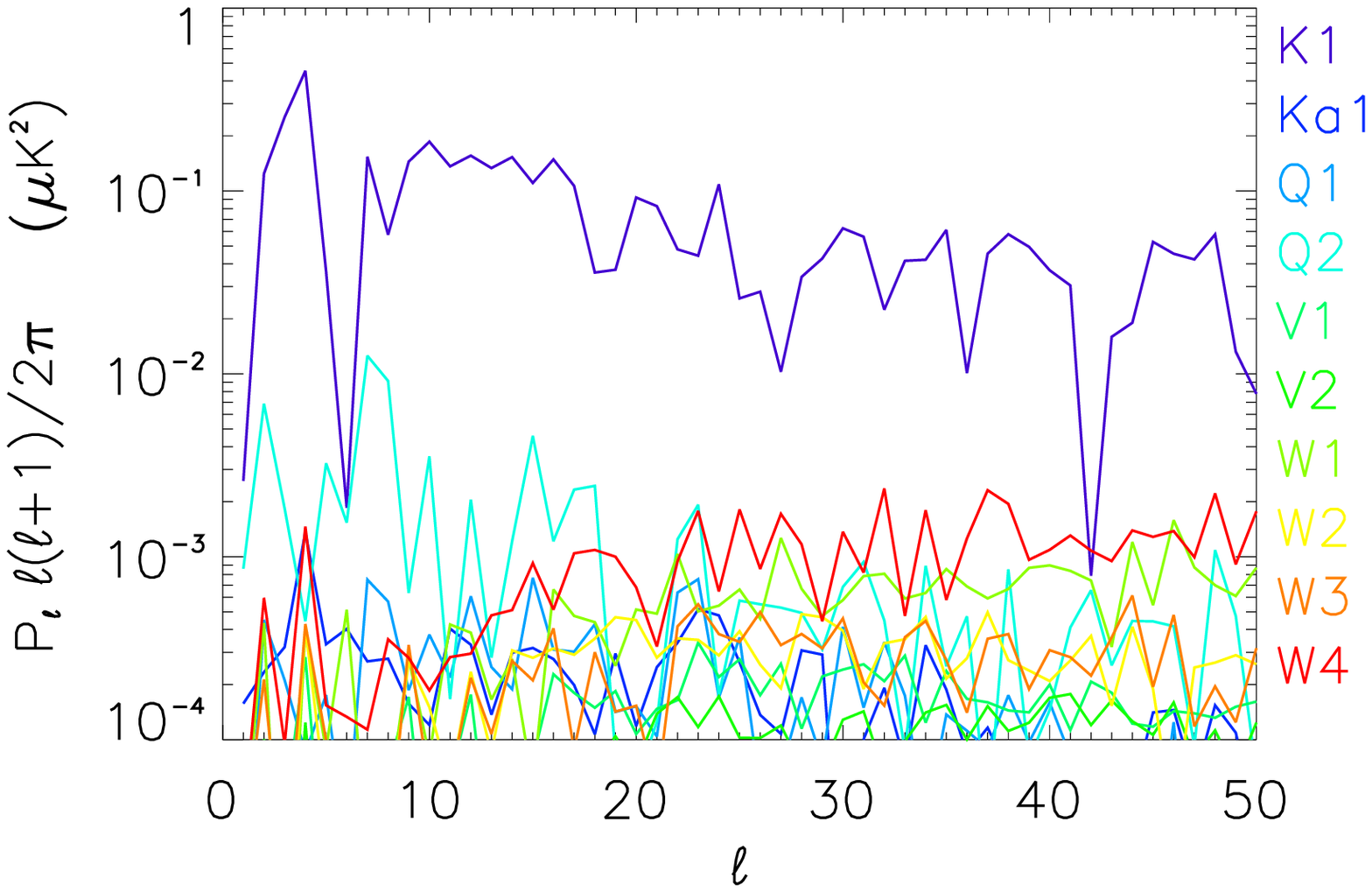}
&
\includegraphics[width=3.0in]{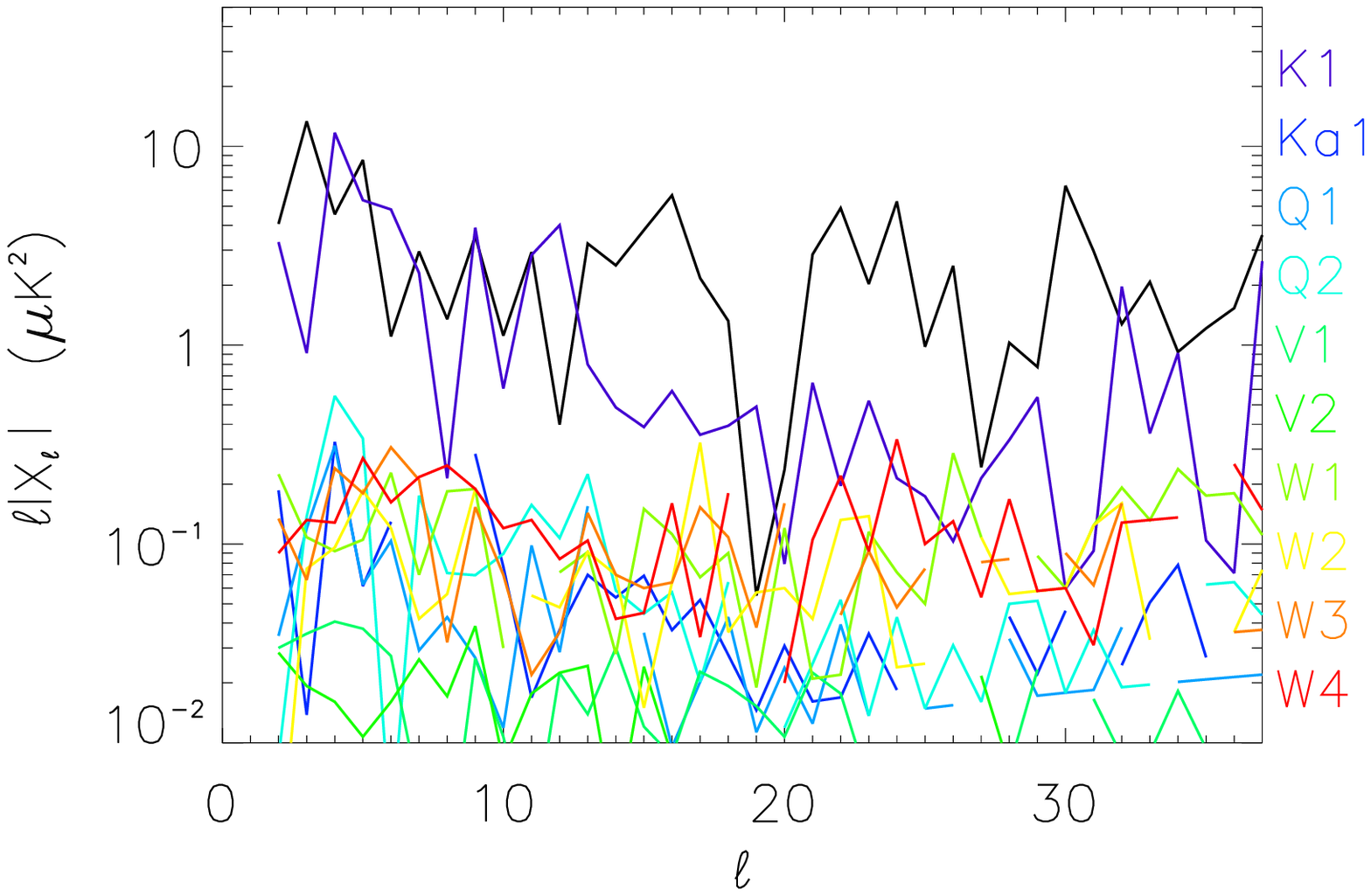}
\\
\caption{Angular power spectra for polarized sidelobe contamination.  These
are spectra for the intensity of the polarized pickup; 
$P_\ell$ is the angular power of $(\delta Q^2 +  \delta U^2)^{1/2}$, 
for each radiometer pair.  Both bandpass mismatch and polarized
foreground pickup are included. This spectrum is not one generally
used for CMB analysis, but serves to show the strength and angular 
dependence of the polarized sidelobe contamination.}
\labfig{pol_spectrum}
&
\caption{The TE angular power spectra for polarized sidelobe
contamination.  These are the largest sidelobe contribution to the 
\WMAP year one TE spectra, from the 
$(\delta Q,\delta U)_{\rm sidelobes}\times(\delta T_{\rm cmb})$.  The
\WMAP reported CMB TE spectrum is shown in black for comparison.
After corrections for Galactic foreground pickup, direct radiometer
bandpass mismatch, and polarized sidelobe pickup (K-band only),
K---W-band data was used to generate the reported TE spectrum at low
$\ell$.  Omitting K-band changes the reported spectrum very little
\protect{\citep{kogut/etal:2003}}.}
\labfig{TE_spectrum}
\end{tabular}
\end{figure*}

\section{Discussion}

For both unpolarized and polarized microwave sky maps, the above
techniques yield measures of the sidelobe contamination in each map
accurate to $\sim$ 30\%.  Most of this uncertainty comes
from the overall calibration error in the sidelobe gains, although a
few percent may be attributed to the approximations in
equations~\eqn{chris_approx}, \eqn{gary_approx}, and \eqn{polavg}, \eqn{poloffset}. 
With the information available, improvements are possible
in a second round of analysis.

The first improvement will be with the calibration of the sidelobe
maps.  With sidelobe maps for all ten horn positions, and
with high-quality sky maps now available, we can return to the
differential time ordered data and extract a best fit calibration for
each gain pattern.  Using the time ordered  $\delta T_X({\cal O})$, we
can bypass the various approximations made in \S
\ref{section:unpolarized}, \S \ref{section:polarized}. 
Using the time ordered data, one can calibrate sidelobe gains directly
from the sky maps, and then correct the data stream directly, prior to
map-making.  \cite{wandelt/gorski:2001} have proposed a mechanism to perform
this calculation efficiently on large data sets.

It appears possible that their method can eliminate sidelobes to below
5\% of their original strength. A similar correction is possible for
the polarized sidelobe pickup.

\subsection{Conclusion}

The systematic signal in the one-year WMAP data induced by sidelobe
pickup of the Galaxy is small.  The sidelobes do not contribute 
strongly to the uncertainties in the CMB anisotropy maps, or in 
the TE angular cross power spectrum. 
Since the sidelobes are broad, smooth features on
the sky, their influence is important only at the largest angular
scales, $\ell < 20$.  Outside of the Galactic region, the magnitude of
overall sidelobe pickup ranges from $60\ \mu \rm K$ in K-band to $1\
\mu \rm K$ in V and W-bands.  Polarized sidelobe pickup is markedly
smaller, ranging from $1 \mu \rm K$ in K-band down to a few nK at
higher frequencies.  

For the first year \WMAP results we restrict ourselves to calculating
sidelobe contamination using the approximations discussed here.  We
subtract a sidelobe signal only from the K-band maps. In future data
sets, we plan to use the raw time ordered data to directly calibrate
the sidelobe maps, and subsequently correct the time-stream data to
remove sidelobe pickup from future maps.

\end{document}